# NEUTRON IRRADIATION OF Mg$^{11}$B$_2$ : FROM THE ENHANCEMENT TO THE SUPPRESSION OF SUPERCONDUCTING PROPERTIES

M. Putti[1], V. Braccini[1], C. Ferdeghini[1], F. Gatti[2], P. Manfrinetti[3], D. Marrè[1], A. Palenzona[3], I. Pallecchi[1], C. Tarantini[1], I. Sheikin[4], H.U. Aebersold[5], E. Lehmann[5]

[1] *INFM-LAMIA, Dipartimento di Fisica, via Dodecaneso 33, 16146 Genova Italy*
[2] *Dipartimento di Fisica, via Dodecaneso 33, 16146 Genova Italy*
[3] *INFM-LAMIA, Dipartimento di Chimica e Chimica Industriale, via Dodecaneso 31, 16146 Genova, Italy*
[4] *GHMFL, MPI-FKF/CNRS, 28 Avenue des Martyrs, BP 166, 38042 Grenoble, France*
[5] *Paul Sherrer Institut, CH-5232 Villagen, Switzerland*

ABSTRACT

In this letter we present the effect of neutron irradiation up to fluences of $3.9 \cdot 10^{19}$ n/cm$^2$ on the superconducting properties of MgB$_2$. In order to obtain a disorder structure homogeneously distributed, the experiment was carried out on bulk samples prepared with the $^{11}$B isotope. Up to fluences of $10^{18}$ n/cm$^2$ the critical temperature is slightly diminished (36 K) and the superconducting properties are significantly improved; the upper critical field is increased from 13.5 T to 20.3 T at 12 K and the irreversibility field is doubled at 5 K. For larger neutron fluences the critical temperature is suppressed down to 12 K and the superconducting properties come out strongly degraded.

Since the discovery of $MgB_2$,[1] special efforts have been made to improve its behavior under magnetic fields by introducing defects through alloying, chemical doping, nanoparticle addition and irradiation. In particular through irradiation it is possible to artificially introduce a controlled amount of disorder: large scale defects produced by irradiation may act as pinning centers leading to an increase of the critical current density $J_c$ and the irreversibility field $H^*$, while point defects decrease the electron mean free path by increasing the residual resistivity $\rho_0$ and $H_{c2}$. Due to the presence of the two gaps, the scenario is complicated in $MgB_2$, where it is predicted that non-magnetic impurity scattering causes pair breaking, as magnetic impurity scattering does in conventional superconductors.[2] Thus, due to the special role of disorder in this material, a careful investigation is required to understand to which extent $H_{c2}$ and $J_c$ can be enhanced without a substantial suppression of $T_c$.

Different irradiation experiments are reported in the literature. Since the beginning, proton irradiation on bulk and single crystals samples has been shown to improve the field dependence of $J_c$ with a significant increase of $H^*$ without strongly suppressing $T_c$.[3,4] Neutron irradiation on bulk samples[5,6] and single crystals[7] produced a slight decrease in $T_c$, an increase in $\rho_0$ and an enhancement of $H_{c2}$, $H^*$ and $J_c$. Only one experiment[8] was carried out with a high neutron dose and led to a strong suppression of $T_c$ that was recovered by annealing. High energy heavy ion irradiation did not produce any substantial effect, but for the increase of $J_c$ over a certain magnetic field range observed on epitaxial thin films[9], while irradiation with α particles increases the resistivity and suppresses the critical temperature.[10]

Until now though, no systematic studies up very high fluences have been performed on bulk material. The creation of homogeneous defect structure in bulk $MgB_2$ samples is not an easy task. In fact, the irradiation with charged particles (protons and ions) have the disvantages of short ranges and varying energy losses along the range. On the other hand neutrons penetrate deeply the materials: irradiations with fast neutrons produce point defects by direct collisions with nuclei, while thermal neutrons give rise to sizable defects mainly by means of capture nuclear reactions. In our case , as discussed in ref.5, neutron irradiation induces defects mainly by neutron capture of $^{10}B$, followed by the emission of an α particle and a $^7Li$ nucleus. Due to the large cross section of this reaction, self-shielding effects prevent a homogeneous neutron penetration over the whole sample. In MgB$_2$ samples produced with natural B (80% of $^{11}$B and 20% of $^{10}$B) the penetration depth comes out to be of the order of 200 μm and in thicker samples the resulting defect structure is therefore highly inhomogeneous with a higher defect density at the surface. To reduce this effect, M. Eisterer *et al.*[5] shielded the samples with cadmium, but the disorder induced by thermal neutrons exceeded by far the damage caused by collisions of fast neutrons with the lattice atoms and in large samples the defect distribution remained rather inhomogeneous.

In this letter we present a systematic study of the effect of neutron irradiation up to very high fluences (3.9·10$^{19}$ n cm$^{-2}$) on the superconducting properties of MgB$_2$. In order to avoid the self shielding effects and to obtain an homogeneous disorder structure the experiment was carried out on bulk samples produced with isotopically pure $^{11}$B.

Very clean samples were prepared by direct synthesis[11] from Mg (99.999% purity) and crystalline isotopically enriched $^{11}$B from Eagle-Picher (99.95% purity). This procedure provides dense and hard cylindrical shaped samples (12 mm diameter) which systematically show optimal critical temperature ($T_C$=39.0-39.2 K), sharp superconducting transition ($\Delta T_C$ ~0.2), low residual resistivity values ($\rho$(40 K)=1-2 μΩcm) and large residual resistivity ratio (RRR=11-15). Two nominally identical samples, cut as parallelepiped bars (~1×1×12 mm$^3$), were irradiated at two different facilities: a TRIGA MARK II type nuclear research reactor (thermal neutron flux density up to 10$^{13}$ cm$^{-2}$s$^{-1}$) at the Laboratory of Applied Nuclear Energy (LENA) of the University of Pavia, and a spallation neutron source SINQ (thermal neutron flux density up to 1.6·10$^{13}$ cm$^{-2}$s$^{-1}$) at the Paul Sherrer Institut (PSI) of Zurich.

X-Ray diffraction (XRD) has been performed on the samples in the standard Bragg-Brentano geometry, and the patterns have been refined through Rietveld analysis to extract the cell parameters. AC electrical resistance measurements were made at a current density of ~1 A/cm$^2$ in a 9 T Quantum Design PPMS. An Oxford 14 T vibrating sample magnetometer (VSM) and a 5.5 T Quantum Design SQUID magnetometer were used for magnetization measurements on LENA and PSI series respectively. The SQUID magnetometer was used for inductive $T_c$ measurements on both series. $H_{c2}(T)$ measurements were performed in the Quantum Design PPMS up to 9 T (LENA series) and in the 20.3T resistive magnet at the GHMFL in Grenoble (PSI series). In table I we report neutron fluence, $T_c$, $\Delta T_c$, $\rho_0$, RRR for the two sample series irradiated at LENA and PSI: L3 and -4 produced by irradiation of L0, and P1 to –5 produced by irradiation of P0. Resistivity vs. temperature curves for all the samples are plotted in figure 1, while in the inset the D.C. susceptibility measurements are shown. The residual resistivity increases monotonously with the neutron fluence, while, at the same time, the RRR decreases. It is worth noticing that the transition remains very sharp both in resistive and inductive measurements: the transition width $\Delta T_c$, as taken between the 10% and the 90% of the resistive transition, remains less than 0.3 K up to fluences of $10^{18}$ n/cm$^2$ and it is less than 1 K also at the largest fluences. This proves that the defects are homogeneously distributed within the samples. In fact, being the $^{10}$B concentration less than 0.5%, the penetration depth of thermal neutrons comes out of the order of 1 cm, much larger than the sample thickness.

Figure 2 shows the behavior of $T_c$, $\rho(40)$, and of the *a* and *c*-axes as a function of the fluence for both the series of samples. $T_c$ monotonously decreases with increasing the neutron fluence; up to $10^{18}$ n/cm$^2$, $T_c$ is only slightly suppressed down to about 36 K, while for larger fluences a significant suppression down to $T_c$=12 K is observed. At the same time, $\rho(40)$ strongly increases by more than two orders of magnitude in the range of fluences we explored, while the temperature dependent term $\rho(300) - \rho(40)$ remains nearly constant (10 - 20 μΩcm). [10]

As from XRD, the samples do not show any secondary phase. The diffraction peaks of irradiated samples remain narrow[8] and the *a*- and *c*-axis are both increased after the introduction of defects, leading to an anisotropic expansion of the crystal lattice. In fact, the c-axis increases more with respect to the a-axis, and the cell volume is expanded up to about 1.8%, in fair agreement with what reported in ref.8. The same way as for the decrease in $T_c$, the increase of the cell parameters is small up to fluences of about $10^{18}$ n/cm$^2$, while it becomes more remarkable for higher fluences.

In Figure 3 $H_{c2}$ vs T for all the irradiated samples is reported. Below fluences of about $10^{18}$ n/cm$^2$, while $T_c$ is only slightly decreased, $H_{c2}$ strongly increases, i.e. from 13.5 T to 20.3 T at 12 K (P3). For higher fluences, both $T_c$ and $H_{c2}$ are strongly depressed.

Critical current density $J_c$ determined from the magnetization hysteresis loops using the appropriate critical state model[12], is shown in Fig. 4 at 5K and 20K. While $J_c$ at 0 T is about constant after irradiation (except for the heavy irradiated samples P4 and P5, where all the superconducting properties are strongly depressed), its magnetic field dependence changes quite strongly. If we define the irreversibility field H* as the field at which $J_c$ falls below 100 A/cm$^2$, we can see a strong increase: at 5 K, where H* doubles passing from 6.8 T (L0) to about 14 (L4) T. This value for H* is comparable to the highest H* values reported in the literature for MgB$_2$ bulk samples. [13]

To analyze this rich and complex phenomenology some hypotheses on the disorder structure should be drawn. M. Zehetmayer et al.[7] have emphasized by transmission electron microscopy the presence of defects, a few nm sized. They could be created by the nuclear reaction products, α and $^7$Li particles, at the end of the recoil range. These defects have a size comparable with the coherence length of MgB$_2$ so that they could be responsible for the increasing of $J_c$ and H*. On the other hand, the increase in resistivity and $H_{c2}$ is related to the decrease in the electron mean free path which is mainly affected by the homogeneously distributed point defects. These defects, which probably cause the increase in the cell volume, can be created along the range of the reaction products. Moreover, due to the small amount of $^{10}$B in our samples, the contribution of fast neutrons which make elastic collisions randomly moving inside the crystal cannot be neglected.

It remains to discuss the strong $T_c$ suppression at high fluences. The degradation of superconductivity induced by disorder was widely investigated in conventional superconductors (see ref. 14 and references therein) while only few studies are available on MgB$_2$.[10,15] NMR data in a heavily neutron irradiated sample[15] showed a reduction of density of states which can contribute to the depression of superconductivity. On the other hand, we observed a strong correlation between the reduction of $T_c$ and the increase of *c*-axis which occur at the same fluence. This correspondence recalls the scenario proposed in ref.16, in which the local distortion in the boron planes due to the atom displacement could play a role in the suppression of the superconductivity in MgB$_2$.

In conclusion, we presented the evolution of the superconducting properties with neutron irradiation. The main results are that large and small scale defects significantly improve the superconducting properties of MgB$_2$ like $J_c$, H* and $H_{c2}$, until the strong suppression of $T_c$ which occurs for fluences larger than $10^{18}$ n/cm$^2$. The overall phenomenology is quite intricate and only a detailed analysis of each property as a function of neutron dose could clarify how electronic structure and different mechanisms, such as inter- and intra-band impurity scattering, electron-phonon coupling, and flux pinnig, are affected by disorder.

TABLE CAPTIONS

Table I: Parameters of the two series of irradiated samples: the critical temperature defined as $T_c=(T_{90\%} + T_{10\%})/2$ and $\Delta T_c = (T_{90\%} - T_{10\%})$, where $T_{90\%}$ and $T_{10\%}$ are estimated at the 90% and 10% of the resistive transition; the resistivity measured at 40 K $\rho(40)$; the residual resistivity ratio defined as RRR=$\rho(300)/\rho(40)$; the crystallographic $a$ and $c$ axes.

FIGURE CAPTIONS

Fig. 1: Resistivity vs. temperature for the virgin and irradiated samples. In the inset, the susceptibility (normalized to -1) is shown as a function of the temperature measured at 1mT after zero field cooling.

Fig.2: Behavior of $T_c$ (a), $\rho(40K)$ (b), a- and c-axes (c-d) with the neutron fluence in the two series of irradiated samples.

Fig. 3: Upper critical field as a function of the temperature as determined from the 90% of the resistive transitions R(T) (PPMS up to 9 T) or R(H) (GHMFL up to 20.3 T)

Fig. 4: Critical current density calculated from M-H loops measured up to 5T and up to 14T for the samples irradiated at LENA and at PSI respectively. The magnetic irreversibility line is determined when $J_c$ falls below 100 A/cm$^2$.

TABLE I

| Samples | Fluence (n/cm$^2$) | $T_c$ (K) | $\Delta T_c$ (K) | RRR | $\rho(40)$ ($\mu\Omega$cm) | a (Å) | c (Å) |
|---|---|---|---|---|---|---|---|
| **L0** | 0 | 39.2 | 0.2 | 14.6 | 0.7 | 3.084 | 3.522 |
| **L3** | 1.0E+17 | 39.2 | 0.2 | 8.2 | 2.0 | 3.086 | 3.526 |
| **L4** | 1.0E+18 | 37.1 | 0.2 | 2.5 | 7.7 | 3.086 | 3.527 |
| **P0** | 0 | 39.2 | 0.2 | 11.1 | 1.6 | 3.084 | 3.519 |
| **P1** | 1.0E+17 | 39.1 | 0.3 | 6.9 | 2.4 | 3.083 | 3.524 |
| **P2** | 6.0E+17 | 37.8 | 0.2 | 3.0 | 6.5 | 3.085 | 3.529 |
| **P3** | 7.6E+17 | 36.1 | 0.3 | 2.0 | 16 | 3.083 | 3.525 |
| **P4** | 1.0E+19 | 24.3 | 0.9 | 1.2 | 64 | 3.088 | 3.549 |
| **P5** | 3.9E+19 | 12.2 | 0.7 | 1.1 | 120 | 3.095 | 3.558 |

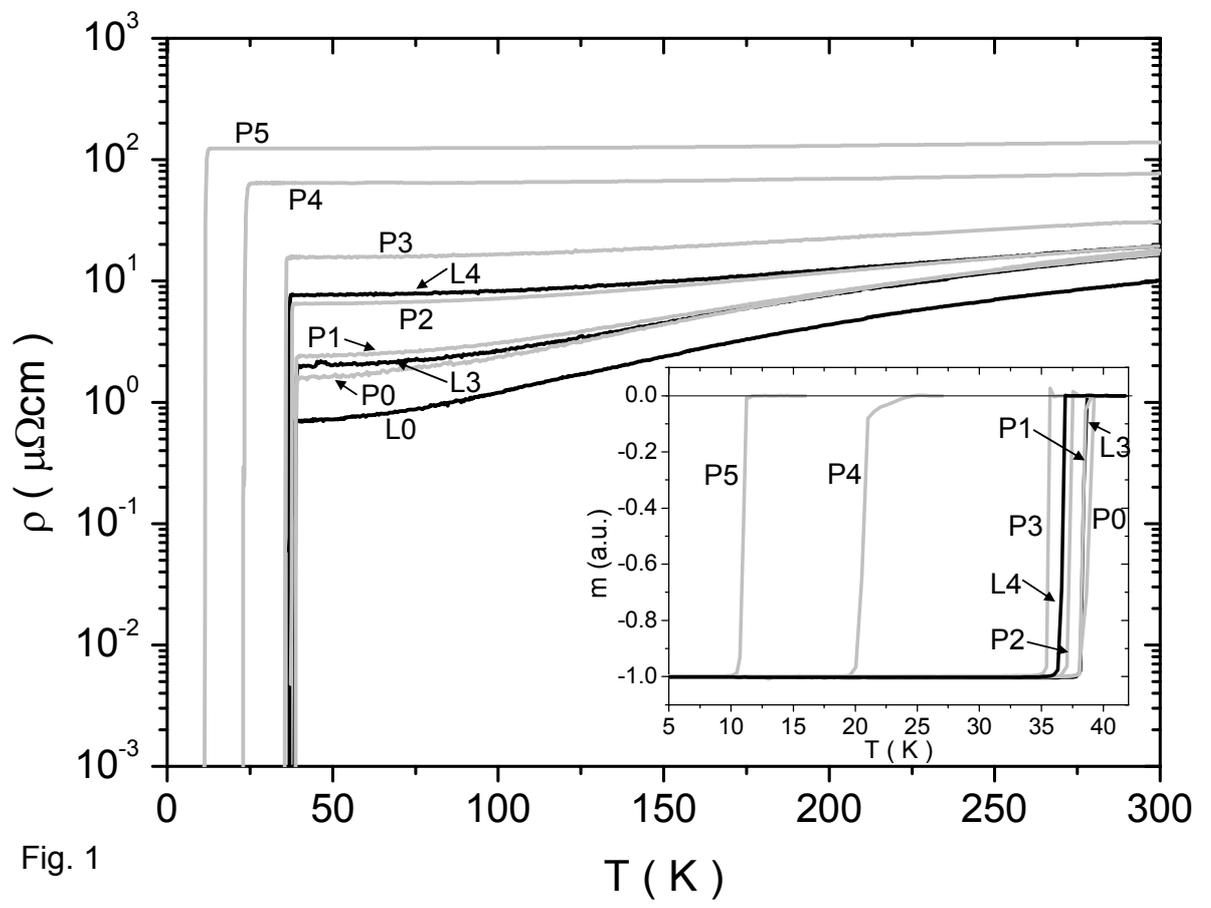

Fig. 1

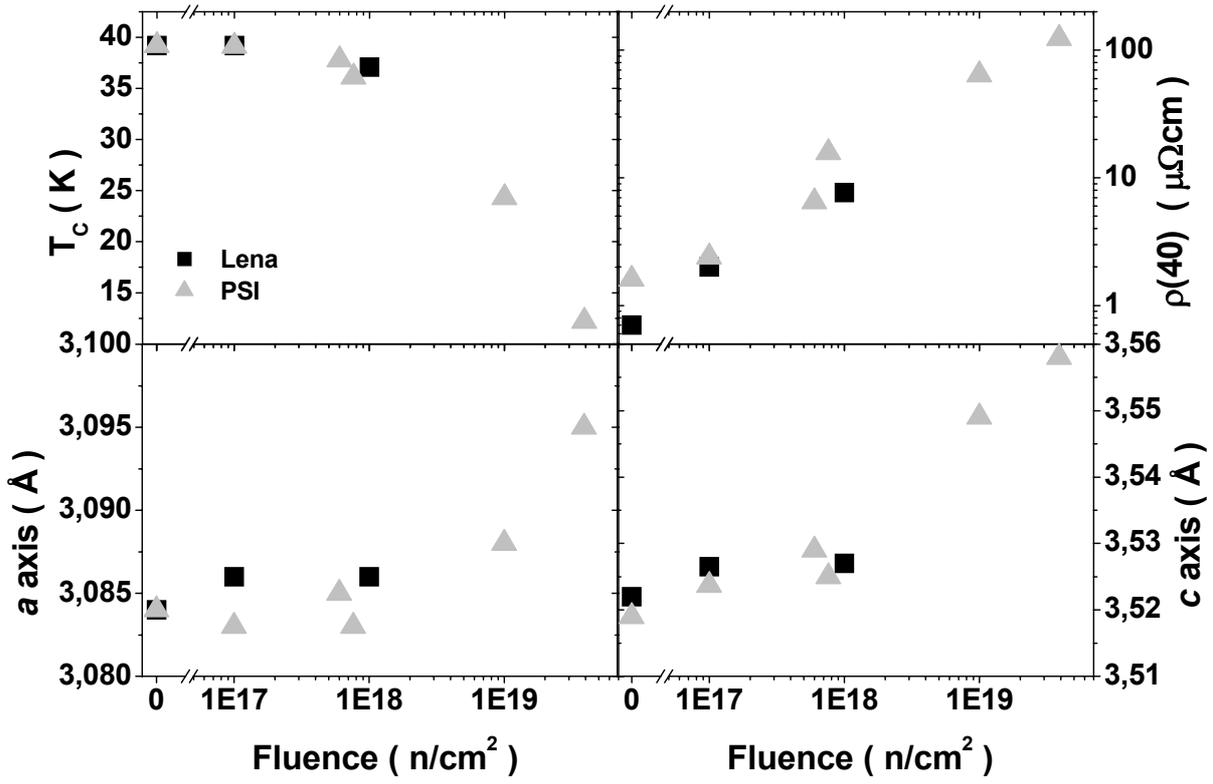

FIG.2

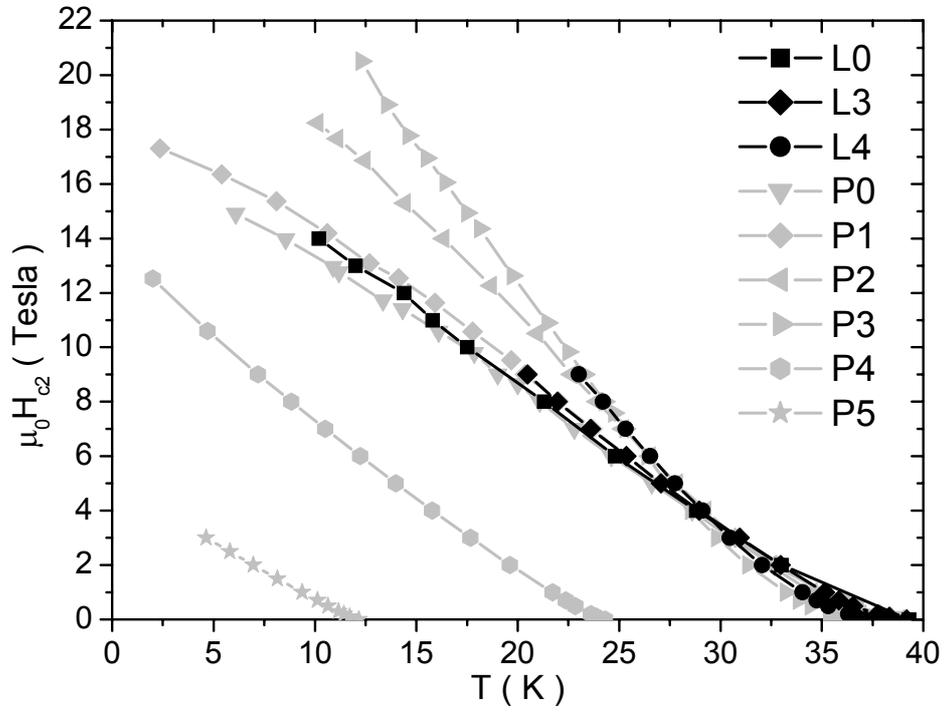

FIG.3

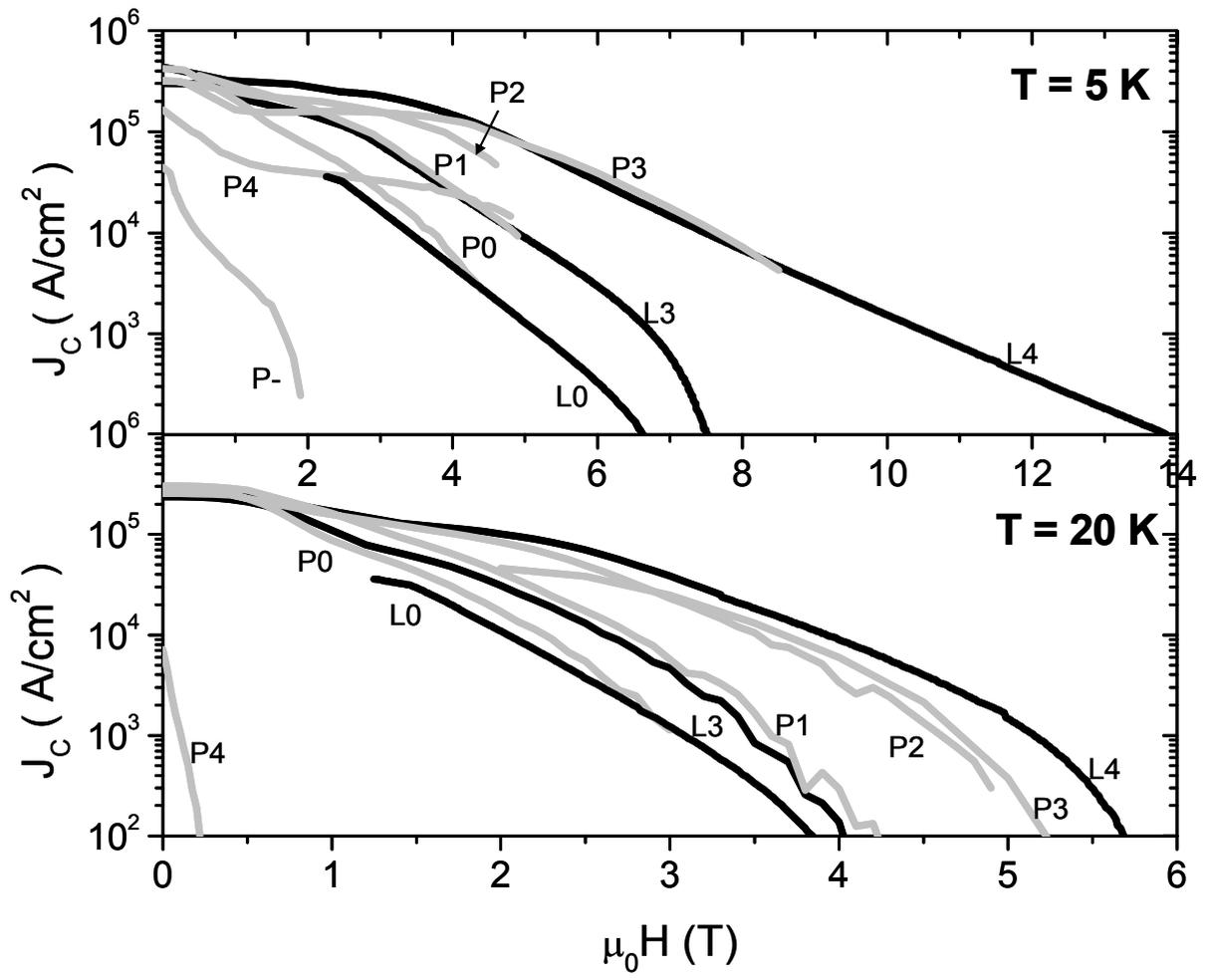

FIG.4